\documentclass[a4paper,nofootinbib,twocolumn,notitlepage, superscriptaddress,longbibliography]{revtex4-1}
\usepackage{amsmath} 					
\usepackage{amssymb}
\usepackage{amsthm}
\usepackage{array}       			
\usepackage{graphicx}  			
\usepackage{xcolor}       			
\usepackage{makeidx}   					
\usepackage[utf8]{inputenc} 			
\usepackage{url}     
\usepackage[english]{babel}
\usepackage{csquotes}
\usepackage{times}
\usepackage{braket}
\usepackage{natbib}
\usepackage[colorlinks=true,breaklinks=true,allcolors=blue]{hyperref}
\usepackage{siunitx}

\newcommand{\cent}{\si{{\degree}}}

\begin{document}	
	\title{Transparent reporting of research-related greenhouse gas emissions\\ through the scientific CO$_{2}$nduct initiative}

\author{Ryan Sweke}
\email{rsweke@gmail.com}
\affiliation{Dahlem Center for Complex Quantum Systems, Freie Universit\"{a}t Berlin, 14195 Berlin, Germany}

\author{Paul Boes}
\affiliation{Dahlem Center for Complex Quantum Systems, Freie Universit\"{a}t Berlin, 14195 Berlin, Germany}

\author{Nelly H.~Y.~Ng}
\affiliation{School of Physical and Mathematical Sciences, Nanyang Technological University, 639673, Singapore}

\author{Carlo Sparaciari}
\affiliation{Department of Physics and Astronomy, University College London, London WC1E 6BT, UK}

\author{Jens Eisert}
\affiliation{Dahlem Center for Complex Quantum Systems, Freie Universit\"{a}t Berlin, 14195 Berlin, Germany}

\affiliation{Helmholtz-Zentrum Berlin f{\"u}r Materialien und Energie, 14109 Berlin, Germany}

\affiliation{Fraunhofer Heinrich Hertz Institute, 10587 Berlin, Germany}

\author{Marcel Goihl}
\affiliation{Dahlem Center for Complex Quantum Systems, Freie Universit\"{a}t Berlin, 14195 Berlin, Germany}
	
\maketitle

{\bf Estimating the greenhouse gas emissions of research-related activities is a critical first step towards the design of mitigation policies and actions. Here we propose and motivate a transparent framework for reporting research-related greenhouse gas emissions, through the inclusion of standardised reporting tables in scientific publications.}
	
\subsection*{Introduction}

\noindent The recent International \emph{Panel on Climate Change} (IPCC) special report on the impacts of global warming to 1.5\cent C above pre-industrial levels has made clear both the scale and urgency of action required, across all sectors of society, to avoid a wide range of severe negative societal and ecological impacts~\cite{ipcc}.  
Science is no exception to this. While scientific research has been extremely important in highlighting the need for action in the first place, it also potentially carries a significant carbon footprint itself.
Given this, many scientists and scientific institutions recognize the responsibility they have in designing and implementing meaningful sustainability strategies.
A first step towards any such strategy is the systematic estimation of research-related greenhouse gas emissions.
However, at most research facilities, frameworks which facilitate this process do not exist. As a result, the requisite data for evidence based decision making and policy design is not available.
To address this, we present a simple framework for the transparent reporting of the greenhouse gas emissions of scientific research. In particular, we propose that authors include a CO$_2$ reporting table in their publications -- such as the example given below in Table \ref{f:example} -- which details the research-related greenhouse gas emissions associated with generating the publication. 
 
 \begin{table}[h]
	\begin{center}
		\begin{tabular}[b]{l c}
			\hline
			\textbf{Numerical simulations} & \\
			Total Kernel Hours [$\mathrm{h}$]& 120000\\
			Thermal Design Power per Kernel [$\mathrm{W}$]& 5.75\\
			Total Energy Consumption of Simulations [$\mathrm{kWh}$] & 1960\\
			Average Emission of CO$_2$ in Germany [$\mathrm{kg/kWh}$]& 0.56\\
			Total CO$_2$-Emission from Numerical Simulations [$\mathrm{kg}$] & 1098\\
			Were the Emissions Offset? & \textbf{Yes}\\
			\textbf{Transportation} & \\
			Total CO$_2$-Emission from Transportation [$\mathrm{kg}$] & 2780 \\
			Were the Emissions Offset? & \textbf{Yes}\\
			\hline
			Total CO$_2$-Emission [$\mathrm{kg}$] & 3878\\
			\hline
		\end{tabular}
		\caption[the caption]%
		{An example of a CO$_2$ reporting table, taken from Sparaciari et al.~\cite{sparaciari2021bounding}. The original caption of the table was as follows: ``Estimated climate footprint of this work. Prototyping is not included in these calculations. Estimations have been calculated using the examples of Scientific CO$_2$nduct \cite{conduct} and are correct to the best of our knowledge." As of today, we would not include the offset line any longer, as the use of offsetting schemes where emissions are mitigated elsewhere is fading. This is due to the adoption within the Paris agreement of national determined contributions, i.e. local emission reduction goals for every country. Instead of allowing entities to generate emission savings in other countries, every country now has to set its own goals, in place of buying offsetting credits.}\label{f:example}
	\end{center}
\end{table}
 
These publication-specific CO$_2$ reporting tables
\begin{enumerate}
    \item are relatively easily produced,
    \item help authors understand and monitor the emissions of their research,
    \item help departments and universities better estimate their carbon footprint, 
    \item help to raise awareness for the carbon footprint of scientific activity, both within the scientific community and beyond it.
\end{enumerate}
The hope is that ultimately such reporting will allow for an assessment of the various contributions to research-related emissions, and therefore facilitate the transformation of scientific research in order to mitigate these. In the following, we will present our proposal in more detail.

\subsection*{Why this framework}

\noindent As mentioned briefly above, we believe there are a variety of reasons which motivate the reporting framework we propose here — namely publication specific greenhouse gas emission reporting tables. Here we elaborate on these reasons.

\begin{enumerate}
    \item \textbf{Ease of production:} There are many organizational levels (e.g., departmental, university-wide) and time-scales (e.g., annually) at which one could attempt to account for research-related emissions. However, we believe that publication-specific reporting tables are particularly convenient, due to the fact that authors of a scientific publication are in a favourable position to estimate the resources which were expended during the course of the research project which lead to the publication. As such, the barriers to the production of such tables are lower than with alternative reporting mechanisms. 
    \item \textbf{Aid universities and departments in better estimating their carbon footprint:} While publication-specific reporting tables can be produced with relative ease, one would ultimately like to be able to obtain estimates of departmental and university-wide carbon footprints. As departments and universities keep track of all affiliated publications, we believe that the presence of publication specific reporting tables -- especially if standardized -- could significantly aid in the process of estimating departmental and university-wide carbon footprints.
    \item \textbf{Assist authors in understanding and monitoring their carbon footprint:} There is currently a growing desire among researchers and policymakers to understand and monitor the greenhouse gas emissions of scientific research. We believe that the process of estimating and tracking the emissions associated with specific research projects will provide an impetus for such monitoring, and will greatly aid researchers in gaining a better understanding of the environmental impact of their research, while simultaneously raising awareness of such issues more broadly. 
    \item \textbf{Raising public awareness in the scientific community and beyond:} While there is currently a growing societal awareness of environmental and sustainability issues, scientists and researchers are often relatively unaware of the environmental impact of their own research. As scientists and researchers are constantly engaging with research publications, we believe that the inclusion of emission reporting tables in publications will help in raising awareness of research-related sustainability issues within the scientific community. Additionally, similarly to the inclusion of funding acknowledgements, such tables will provide transparency to the broader public, who are ultimately the main contributors to the funding of scientific research.
\end{enumerate}

\subsection*{Creating a greenhouse gas emission reporting table}

Having provided a motivation for our suggested reporting framework, in this section we discuss the process of creating a publication specific CO$_2$ reporting table. While there are well established carbon reporting schemes whose methodology is publicly available -- such as the greenhouse gas protocol~\cite{ghg} for product life-cycles -- these schemes typically focus on the carbon footprint of either whole companies or the full life cycle of a product, and as such require some adaptation for estimating emissions of a research project.  Given this, before creating the table, we first need to collect the necessary data to estimate the carbon footprint of the project in question. We opt for a very simple guideline: doing the best we can. More specifically, we suggest trying to capture as many sources as exactly as possible, while at the same time acknowledging that there are limits to what can be reasonably estimated or included. Ultimately, we would like to encourage reporting even weak estimates as a first step, provided they are acknowledged as such.
Let us now give a simple step-by-step manual on how to estimate the carbon footprint for a given project.
\begin{enumerate}
    \item Define the scope of the analysis. This means fixing
    the time frame to analyse, which parts of your work
    contributed to this specific project or which locations are 
    relevant to the analysis.
    \item Collect the relevant emissions sources. With respect to the previously defined scope, list the emission sources that contributed
    to the project. These can be for example flights to a conference,
    where the project was presented or experiments which contributed to the results.
    \item Estimate the footprint of the carbon sources. For each of the 
    sources, calculate the emissions. In the ideal case, there
    is direct data for the calculation, such as the fuel used to drive
    to a conference, or a power meter for the computer used. In many cases, however, either default values or estimates have to be used.
    The underlying assumptions should be justifiable.
\end{enumerate}
The process has to be carried out iteratively. For example, it may
happen that in
the final step there is no data available to estimate a source or
that estimates show that a specific source is negligible compared
to other emissions, making it necessary to revert to the second step;
or during the analysis it may transpire that the work will be presented
at a conference, making the authors re-evaluate the scope.

Having completed step three of the list above, you are now in
position to present the carbon footprint as part of the resulting
publication. For concreteness, we will walk you through the creation of the reporting table in Table~\ref{f:example}, which was already included in the work of Sparaciari et. al~\cite{sparaciari2021bounding}, a recent theoretical physics publication which combined both
analytical and numerical work.
For the scope, the authors decided to include a specific conference
visit where they met and outlined the original idea 
for the project, as well as the numerical 
simulations which produced the results shown in the publication.
This meant that all the prototyping prior to the final results was
excluded. While prototyping certainly did contribute to the emissions of the project, data which would have allowed for meaningful estimations was not collected during this phase of the research project. We note that this exclusion was mentioned explicitly in the caption of the reporting table. Other factors such as commuting and heating/cooling of the
office were neglected, as first estimates showed that the corresponding
emissions were small compared to the other sources.

For the conference visit, the predominant source were the emissions
from flights. These were estimated using free online tools for 
converting flight data to emission data. The more information about
the flight is available, e.g. the airline used, the better the 
estimate.
For the numerical calculations, the authors used a supercomputer
at their department. The power consumption of the kernels in the
supercomputer were made publicly available by the manufacturer. Knowing 
the simulation runtime, the authors then estimated the power consumption
of the numerical results. At this point, a default value (see for example
the data from the European Energy Agency \cite{eea}) for the 
carbon emissions per unit electricity was employed to obtain
the final emission estimate for the numerical simulations.
The results have were collected into the table shown in Table~\ref{f:example}, which was then included in the 
published work \cite{sparaciari2021bounding}.

As already mentioned briefly, we note that the example above is incomplete in two ways. Firstly, not all the sources that contributed to the project were part of the final report. Some were in fact purposefully neglected, as first estimates showed that they would be irrelevant. For others, there was no
data available. In the case of the numerical prototyping,
data would have been available had the authors been aware of the inventory prior to starting the project. Once again, however, we stress that our approach is to do the best we can, provided the shortcomings of the analysis are explicitly acknowledged.
Secondly, we stress that the above example is for a work in theoretical physics, performed by a small collaboration, for which numerical simulations and flights were the main sources of emissions. However, for research in other fields, or for larger collaborations, there may be different or additional sources of emissions. In light of this, we give here an incomplete 
overview of additional possible emission sources, which may be relevant for research in different disciplines. 

\begin{itemize}
    \item {\bf Experiments:} Ranging from power consumption of
    measurement devices to fridges as well as potential direct
    release of greenhouse gases, e.g. from chemical reactions.
    \item {\bf Infrastructure:} Heating, cooling and lighting of the 
    research work space. Usually a minor contributor to project
    emissions and can be difficult to assess.
    \item {\bf Commuting:} Contribution depends on the means of 
    transportation.
    \item {\bf Procurement:} Potentially very relevant for the footprint,
    but at the same time very difficult to assess. Asking for a product
    footprint when buying equipment can help to increase the awareness
    also at the level of equipment suppliers.
    \item {\bf Waste disposal:} Especially relevant for laboratories producing hazardous waste.
    \item {\bf Conference-related emissions:} Such as lodging, diet
    and infrastructure. Usually a minor contributor to the 
    emissions of a project and can be difficult to assess.
\end{itemize}

\subsection*{The Scientific CO$_2$nduct community}

At the moment, there exists a small but increasing group of motivated researchers adopting the framework we propose here. It is clear that this initiative would benefit from network effects and increased adoption quite heavily beyond the obvious aspect of gaining traction. For example, the work required to compose the proposed tables and include them in publications could be significantly lowered if researchers had access to a repository of existing tables, templates and model calculations as well as a community of other researchers that can assist them in the process. This is why we have launched an open source website, \href{https://scientific-conduct.github.io/}{https://scientific-conduct.github.io/}, that provides a simple platform for sharing and accessing these resources. The current focus for the site is to explain the initiative and collect publications with the included table, but depending on demand additional features can, and should, be added. For the sake of concreteness, let us finish with a list of ways in which the interested reader could contribute:
\begin{itemize}
    \item Include a table of carbon emissions in your next work (a simple LaTeX template can be found on the website). We encourage you to include any additional sources or features which may be relevant. For example, for future reporting purposes it may be useful to include the emissions per institution involved in the research project, if this can feasibly be calculated.
    \item If you add the table to your manuscript and upload it to a preprint platform such as the arXiv, let the whole community know about it; in the comments field, add ``1 CO2 reporting table'' alongside the number of pages and figures present in your pre-print.
    \item When presenting the results of your work at conferences and seminars, dedicate the last slide of your talk to the carbon emissions generated by the project. Additionally,  refer the audience to the Scientific Co$_2$nduct website for further information on the reporting framework. The authors of this article have done so on several occasions, always to an overwhelmingly positive response from attendees.
    \item Please consider adding this work to the list of examples on the website (preferably via pull request, otherwise via email to any one of the authors).
    \item Consider documenting the calculation for your table and also submitting it to the website. If not the numbers themselves, your methodology will probably be useful for others trying to get started. 
    \item The aim for the website is to become a repository that makes the generation of the proposed tables easy and comparable. If you want to get involved in supporting this initiative beyond using the tables for your own research, we would be very happy to hear from you.
\end{itemize}

\subsection*{Acknowledgements}

\noindent We warmly thank Diana Born for feedback on the draft, as well as the Eisert group at the Freie Universit\"{a}t Berlin for implementing and further developing this framework. R.~S.~and J.~E.~have been supported by the BMWK (PlanQK).

%\bibliography{literature.bib}

\begin{thebibliography}{5}%
\makeatletter
\providecommand \@ifxundefined [1]{%
 \@ifx{#1\undefined}
}%
\providecommand \@ifnum [1]{%
 \ifnum #1\expandafter \@firstoftwo
 \else \expandafter \@secondoftwo
 \fi
}%
\providecommand \@ifx [1]{%
 \ifx #1\expandafter \@firstoftwo
 \else \expandafter \@secondoftwo
 \fi
}%
\providecommand \natexlab [1]{#1}%
\providecommand \enquote  [1]{``#1''}%
\providecommand \bibnamefont  [1]{#1}%
\providecommand \bibfnamefont [1]{#1}%
\providecommand \citenamefont [1]{#1}%
\providecommand \href@noop [0]{\@secondoftwo}%
\providecommand \href [0]{\begingroup \@sanitize@url \@href}%
\providecommand \@href[1]{\@@startlink{#1}\@@href}%
\providecommand \@@href[1]{\endgroup#1\@@endlink}%
\providecommand \@sanitize@url [0]{\catcode `\\12\catcode `\$12\catcode
  `\&12\catcode `\#12\catcode `\^12\catcode `\_12\catcode `\%12\relax}%
\providecommand \@@startlink[1]{}%
\providecommand \@@endlink[0]{}%
\providecommand \url  [0]{\begingroup\@sanitize@url \@url }%
\providecommand \@url [1]{\endgroup\@href {#1}{\urlprefix }}%
\providecommand \urlprefix  [0]{URL }%
\providecommand \Eprint [0]{\href }%
\providecommand \doibase [0]{http://dx.doi.org/}%
\providecommand \selectlanguage [0]{\@gobble}%
\providecommand \bibinfo  [0]{\@secondoftwo}%
\providecommand \bibfield  [0]{\@secondoftwo}%
\providecommand \translation [1]{[#1]}%
\providecommand \BibitemOpen [0]{}%
\providecommand \bibitemStop [0]{}%
\providecommand \bibitemNoStop [0]{.\EOS\space}%
\providecommand \EOS [0]{\spacefactor3000\relax}%
\providecommand \BibitemShut  [1]{\csname bibitem#1\endcsname}%
\let\auto@bib@innerbib\@empty
%</preamble>
\bibitem [{\citenamefont {Allen}\ \emph {et~al.}(2018)\citenamefont {Allen},
  \citenamefont {Dube}, \citenamefont {Solecki}, \citenamefont
  {Arag{\'o}n-Durand}, \citenamefont {Cramer}, \citenamefont {Humphreys},
  \citenamefont {Kainuma}, \citenamefont {Kala}, \citenamefont {Mahowald},
  \citenamefont {Mulugetta} \emph {et~al.}}]{ipcc}%
  \BibitemOpen
  \bibfield  {author} {\bibinfo {author} {\bibfnamefont {M.}~\bibnamefont
  {Allen}}, \bibinfo {author} {\bibfnamefont {O.~P.}\ \bibnamefont {Dube}},
  \bibinfo {author} {\bibfnamefont {W.}~\bibnamefont {Solecki}}, \bibinfo
  {author} {\bibfnamefont {F.}~\bibnamefont {Arag{\'o}n-Durand}}, \bibinfo
  {author} {\bibfnamefont {W.}~\bibnamefont {Cramer}}, \bibinfo {author}
  {\bibfnamefont {S.}~\bibnamefont {Humphreys}}, \bibinfo {author}
  {\bibfnamefont {M.}~\bibnamefont {Kainuma}}, \bibinfo {author} {\bibfnamefont
  {J.}~\bibnamefont {Kala}}, \bibinfo {author} {\bibfnamefont {N.}~\bibnamefont
  {Mahowald}}, \bibinfo {author} {\bibfnamefont {Y.}~\bibnamefont {Mulugetta}},
   \emph {et~al.},\ }\bibfield  {title} {\enquote {\bibinfo {title} {Chapter 1:
  Framing and context},}\ }in\ \href@noop {} {\emph {\bibinfo {booktitle}
  {Global Warming of 1.5°C. An IPCC Special Report on the impacts of global
  warming of 1.5°C above pre-industrial levels and related global greenhouse
  gas emission pathways, in the context of strengthening the global response to
  the threat of climate change, sustainable development, and efforts to
  eradicate poverty}}},\ \bibinfo {editor} {edited by\ \bibinfo {editor}
  {\bibfnamefont {V.}~\bibnamefont {Masson-Delmotte}}, \bibinfo {editor}
  {\bibfnamefont {H.-O.~Pörtner}\ \bibnamefont {P.~Zhai}}, \bibinfo {editor}
  {\bibfnamefont {D.}~\bibnamefont {Roberts}}, \bibinfo {editor} {\bibfnamefont
  {J.}~\bibnamefont {Skea}}, \bibinfo {editor} {\bibfnamefont {P.~R.}\
  \bibnamefont {Shukla}}, \bibinfo {editor} {\bibfnamefont {A.}~\bibnamefont
  {Pirani}}, \bibinfo {editor} {\bibfnamefont {W.}~\bibnamefont
  {Moufouma-Okia}}, \bibinfo {editor} {\bibfnamefont {C.}~\bibnamefont
  {Péan}}, \bibinfo {editor} {\bibfnamefont {R.}~\bibnamefont {Pidcock}},
  \bibinfo {editor} {\bibfnamefont {S.}~\bibnamefont {Connors}}, \bibinfo
  {editor} {\bibfnamefont {J.~B.~R.}\ \bibnamefont {Matthews}}, \bibinfo
  {editor} {\bibfnamefont {Y.}~\bibnamefont {Chen}}, \bibinfo {editor}
  {\bibfnamefont {X.}~\bibnamefont {Zhou}}, \bibinfo {editor} {\bibfnamefont
  {M.I.}\ \bibnamefont {Gomis}}, \bibinfo {editor} {\bibfnamefont {E.~Lonnoy}\
  \bibnamefont {amd T.~Maycock}}, \bibinfo {editor} {\bibfnamefont
  {M.}~\bibnamefont {Tignor}}, \ and\ \bibinfo {editor} {\bibfnamefont
  {T.}~\bibnamefont {Waterfield}}}\ (\bibinfo {year} {2018})\BibitemShut
  {NoStop}%
\bibitem [{\citenamefont {Sparaciari}\ \emph {et~al.}(2021)\citenamefont
  {Sparaciari}, \citenamefont {Goihl}, \citenamefont {Boes}, \citenamefont
  {Eisert},\ and\ \citenamefont {Ng}}]{sparaciari2021bounding}%
  \BibitemOpen
  \bibfield  {author} {\bibinfo {author} {\bibfnamefont {C.}~\bibnamefont
  {Sparaciari}}, \bibinfo {author} {\bibfnamefont {M.}~\bibnamefont {Goihl}},
  \bibinfo {author} {\bibfnamefont {P.}~\bibnamefont {Boes}}, \bibinfo {author}
  {\bibfnamefont {J.}~\bibnamefont {Eisert}}, \ and\ \bibinfo {author}
  {\bibfnamefont {N.~H.~Y}\ \bibnamefont {Ng}},\ }\bibfield  {title} {\enquote
  {\bibinfo {title} {Bounding the resources for thermalizing many-body
  localized systems},}\ }\href@noop {} {\bibfield  {journal} {\bibinfo
  {journal} {Comm. Phys.}\ }\textbf {\bibinfo {volume} {4}},\ \bibinfo {pages}
  {1--8} (\bibinfo {year} {2021})}\BibitemShut {NoStop}%
\bibitem [{con()}]{conduct}%
  \BibitemOpen
  \href {https://scientific-conduct.github.io} {\enquote {\bibinfo {title}
  {Scientific co2nduct},}\ }\bibinfo {howpublished} {online}\BibitemShut
  {NoStop}%
\bibitem [{ghg()}]{ghg}%
  \BibitemOpen
  \href {https://ghgprotocol.org/product-standard} {\enquote {\bibinfo {title}
  {Greenhouse gas protocol},}\ }\bibinfo {howpublished} {online}\BibitemShut
  {NoStop}%
\bibitem [{eea()}]{eea}%
  \BibitemOpen
  \href
  {https://www.eea.europa.eu/data-and-maps/daviz/co2-emission-intensity-8#tab-googlechartid_googlechartid_googlechartid_googlechartid_chart_11111}
  {\enquote {\bibinfo {title} {European energy agency},}\ }\bibinfo
  {howpublished} {online}\BibitemShut {NoStop}%
\end{thebibliography}

\end{document}